# Coexistence of Interfacial Stress and Charge Transfer in Graphene Oxide based Magnetic Nanocomposites

**Amodini Mishra[1], Vikash Kumar Singh[2] and Tanuja Mohanty[1*]**

[1]School of Physical Sciences, Jawaharlal Nehru University, New Delhi, India, 110067

[2]Solid State Physics Laboratory, Timarpur, New Delhi, India, 110054

**Abstract** In this paper, we establish the existence of both compressive stress and charge transfer process in hydrothermally synthesized cobalt ferrite-graphene oxide ($CoFe_2O_4$/GO) nanocomposites. Transmission electron microscopy (TEM) results reveal the decoration of $CoFe_2O_4$ nanoparticles on GO sheets. Magnetic response of nanocomposites was confirme from superconducting quantum interference device (SQUID) magnetometer measurement. Optical properties of these nanocomposites were investigated by Raman spectroscopy. Interfacial compressive stress involved in this system is evaluated from observed blue shift of characteristic G peak of graphene oxide. Increase in full width half-maximum ( FWHM) as well as up shift in D and G peaks are clear indicator of involvement of charge transfer process between GO sheets and dispersed magnetic nanoparticles. The effect of charge transfer process is quantified in terms of shifting of Fermi level of these nanocomposites. This is evaluated from variation in contact surface potential difference (CPD) using Scanning Kelvin probe microscopy (SKPM). XRD spectra of $CoFe_2O_4$/GO confirm the polycrystalline nature of $CoFe_2O_4$ nanoparticles. Lattice strain estimated from XRD peaks are correlated to the observed Raman shift.

**Keywords**: Graphene Oxide, Stress, Raman spectra, charge transfer and Fermi level.

## Introduction

Graphene oxide (GO), a derivative of graphene, is an excellent two dimensional (2D) flat sheet of hexagonally bonded $sp^2$ hybrid carbon atoms on which oxygen functional groups (such as hydroxyl, epoxide, carboxyl and carbonyl and hydroxyl groups) are covalently linked on their basal planes. [1-6] These functional groups attached to the negatively charged GO sheets can act as host for the positively charged ions of nanoparticles (NPs) leading to formation of nanocomposites. These nanocomposites exhibit unique properties as compared to individual components and therefore find a wide range of applications in surface-enhanced Raman scattering, sensors, catalysis and optoelectronic devices.[7-10] Nanocomposites synthesized by hydrothermal route are found to be in the form of dispersion of nucleated nanoparticles on GO sheets. It is a well known fact that, magnetic nanoparticles (MNPs) have potential applications in the field of energy storage devices, MRI and magnetic field driven drug delivery.[11-12] Particularly, Cobalt ferrite ($CoFe_2O_4$) NPs have found lots of applications in the field of catalysis and magnetism based nano devices due to their remarkable chemical and mechanical stability, magnetic behavior, low toxicity and biocompatibility in physiological environments. [13-15] However, pristine $CoFe_2O_4$ NPs suffers from irreversible aggregation and settling due to strong dipole-dipole interaction which can be overcome by employing viscoelastic carrier or surfactant. Particularly two dimensional (2D) planar structures like GO sheets are expected to solve this sedimentation problem by acting as a carrier which enhances the properties of $CoFe_2O_4$ NPs in nanocomposites forms and expands its application possibilities. These $CoFe_2O_4$/GO nanocomposite are considered as one of the most promising electrode materials due to its high abundance, strong magnetic properties, low toxicity as well as cost effectiveness.[16-20]

In past few years, although a lot of interest has been developed by different scientific groups on the magnetic properties and applications of the $CoFe_2O_4$/GO nanocomposite materials, yet, studies on surface electronic, interfacial stress and charge transfer phenomenon of these materials are still unexplored. This research report emphasizes on the existence of charge transfer as well as compressive stress in $CoFe_2O_4$/GO magnetic nanocomposite which are confirmed from Raman spectroscopy, XRD and scanning Kelvin probe measurement. The surface electronic property, particularly shifting of Fermi surface is monitored using scanning Kelvin probe microscopy, where, it is measured in terms of variation in the contact surface potential difference (CPD). The morphology and structure of nanocomposite were examined using scanning electron microscopy and transmission electron microscopy. Their magnetic response was studied using SQUID.

**Experimental Methods**

**Materials**

Graphite flakes (99.8 %, 325 mesh) was purchased from Alfa Aesar.Hydrazine hydrate ($N_2H_4$) and sulfuric acid ($H_2SO_4$, 95%) are procured from Sigma–Aldrich.Potassium permanganate ($KMnO_4$), sodium nitrate, ($NaNO_3$) hydrogen peroxide ($H_2O_2$), ethanol, hydrochloric acid (HCl), Cobalt(II) nitrate hexahydrate [$Co(NO_3)_2 \cdot 6H_2O$], Iron(III) nitrate nonahydrate [$Fe(NO_3)_3 \cdot 9H_2O$], ammonium hydroxide ($NH_4OH$) and double distilled water were purchased from Merck. All the chemicals are used for experiment without further purification.

**Synthesis of graphene oxide sheets**

Graphene oxide (GO) was synthesized from graphite flakes using modified Hummers' method.[21]

**Synthesis of CoFe$_2$O$_4$/GO nanocomposite**

In the first step, **g**raphene oxide (GO) was synthesized from graphite flakes using modified Hummers' method.[21] Graphene oxide based cobalt ferrite nanocomposite (CoFe$_2$O$_4$/GO) was prepared by the hydrothermal method using Cobalt(II) nitrate hexahydrate [$Co(NO_3)_2 \cdot 6H_2O$], Iron(III) nitrate nonahydrate [$Fe(NO_3)_3 \cdot 9H_2O$].[16] For this purpose, firstly 0.25 g of graphene oxide powder was added in 80 mL of ethanol and completely dispersed by ultrasonication for 60 min. In the second step, 0.3 g of $Co(NO_3)_2 \cdot 6H_2O$ and 0.9 g of $Fe(NO_3)_3 \cdot 9H_2O$ were dissolved in 50 mL of ethanol followed by stirring for 3h. The solution was mixed dropwise into the GO suspension with continuous stirring for 5 h. After that, 4.3 g of sodium acetate ($CH_3COONa$) was added into the mixture under continuous stirring. After agitation for 8 h, the mixture solution was transferred to a Teflon-line autoclave. The autoclave was heated under oven at 200 °C for 24 h and then cooled down to room temperature. The solid product was separated by centrifugation and washed thoroughly with water and absolute ethanol to remove impurities. Finally, the product was dried in an oven at 50 °C for a full night. The final product was labeled as graphene oxide based cobalt ferrite nanocomposite (CoFe$_2$O$_4$/GO). The steps involved during synthesis process of CoFe$_2$O$_4$/GO nanocomposite are schematically illustrated in Fig.1.

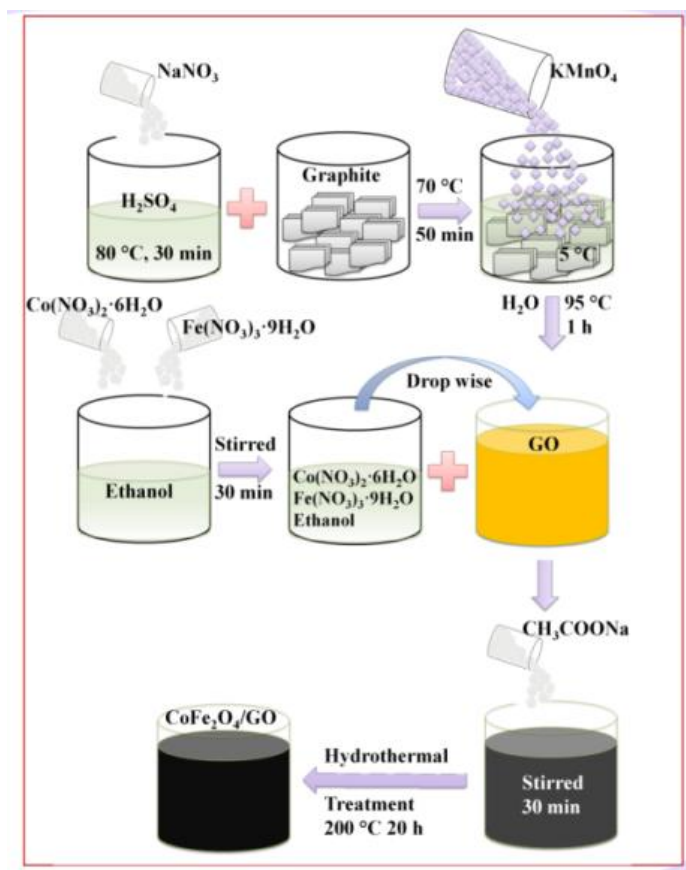

**Figure 1** Schematic representation of steps involved in the synthesis of $CoFe_2O_4$/GO.
nanocomposite.

**Characterization**

The surface morphology of GO and $CoFe_2O_4$/GO was investigated by scanning electron microscopy (SEM) (Care-Zeiss EVO-40, working voltage 20 kV, Germany). The elemental identification of $CoFe_2O_4$/GO nanocomposite was confirmed from energy dispersive X-ray analysis (EDAX). EDAX measurement was carried out using a (Zeiss EVO ED15) microscope coupled with an (Oxford-X-MaxN) EDX detector. The magnetic properties of this nanocomposite were investigated at room temperature using a Quantum Design MPMS-7 SQUID magnetometer. From the magnetization versus applied field plot (M-H), the saturation magnetization ($M_s$), coercivity ($H_c$) and remanence magnetization ($M_r$) was measured. For structural analysis, transmission electron microscopy study was carried out by 200 kV TEM, JEOL 2100F, Japan. X-ray diffraction (XRD) spectra of $CoFe_2O_4$/GO nanocomposite samples were recorded using an X-ray diffractometer (Panalytical 2550-PCX-raydiffractometer). XRD data were collected using Cu-K$_\alpha$ ($\lambda$ = 0.154 nm) radiation with 2θ ranging from 10° to 70° at scanning rate 3° min$^{-1}$. Optical properties of GO and $CoFe_2O_4$/GO nanocomposite were investigated using Raman spectroscopy (HORIBA Xplora) having green laser ($\lambda$= 514 nm) excitation with a laser spot size 1 μm. The effect of $CoFe_2O_4$ nanoparticles decoration on the Fermi energy level of GO sheets is monitored by scanning Kelvin probe microscopy (SKPM, KP Technology, UK).

**Results and discussion**

**SEM and EDAX Studies**

SEM image and EDAX spectra analysis of GO and $CoFe_2O_4$/GO nanocomposite thin film was carried out for observation of the surface morphology and identification of the elements present in the GO and $CoFe_2O_4$/GO nanocomposite as shown in Fig. 2. In the SEM image of GO, crumpled sheets like structures are observed, while in the case of $CoFe_2O_4$/GO nanocomposite, clustering of $CoFe_2O_4$ nanoparticles on GO sheet is found.

EDAX spectrum of GO sheets confirms the presence of C, O and Si elements and that of $CoFe_2O_4$/GO nanocomposite indicates the prominent presence of C, O, Co, Fe, and Si elements. Additional Si peak arising in both the cases are from the silicon substrate. The peaks in the EDAX pattern were perfectly assigned to the elements present in $CoFe_2O_4$/GO composites EDAX spectra of GO and $CoFe_2O_4$/GO nanocomposite clearly signify the high purity in chemical composition of $CoFe_2O_4$/GO nanocomposite.

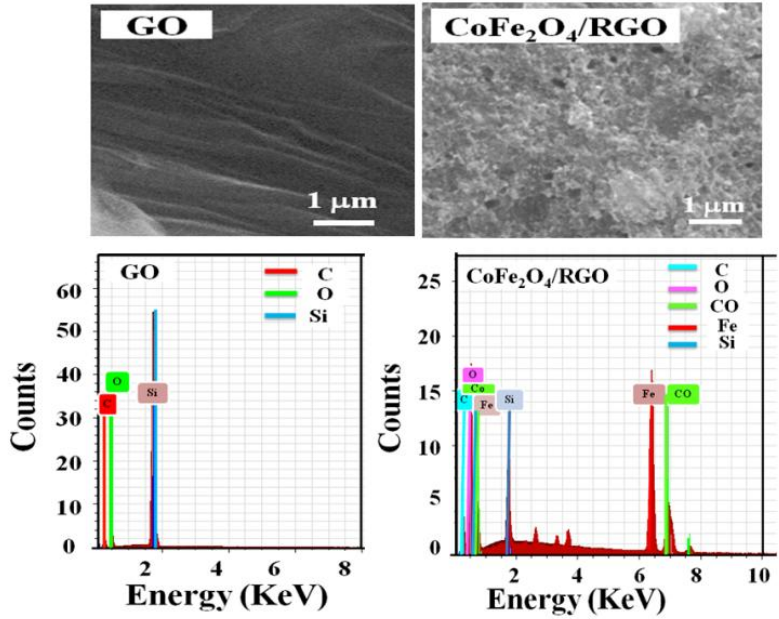

**Figure 2** SEM image and EDAX spectrum of GO and CoFe$_2$O$_4$/GO nanocomposite.

**SQUID Measurement**

Magnetic properties of this nanocomposite were investigated using a SQUID. The M-H loop for the CoFe$_2$O$_4$/GO nanocomposite at 300 K (room temperature) is shown in Fig. 3 (a). At room temperature, the value of saturation magnetization 'M$_s$' comes out to be 75.37 emu/g which is lower than that of corresponding pure bulk CoFe$_2$O$_4$ (94 emu/g). The remanence magnitude 'M$_r$' extracted from the hysteresis loop at the intersections of the loop (shown in the inset) with the vertical magnetization axis is found to be 20.05 emu/g. The coercivity H$_c$ obtained from hysteresis loop is 0.41 kOe for CoFe$_2$O$_4$/GO nanocomposite[22,23] which is quite low thus indicating its soft magnetic nature. These CoFe$_2$O$_4$/GO nanocomposite exhibit a ferromagnetic behavior having small remnant magnetization and coercivity, which is desirable for many practical applications that required strong magnetic signals at small applied magnetic fields.

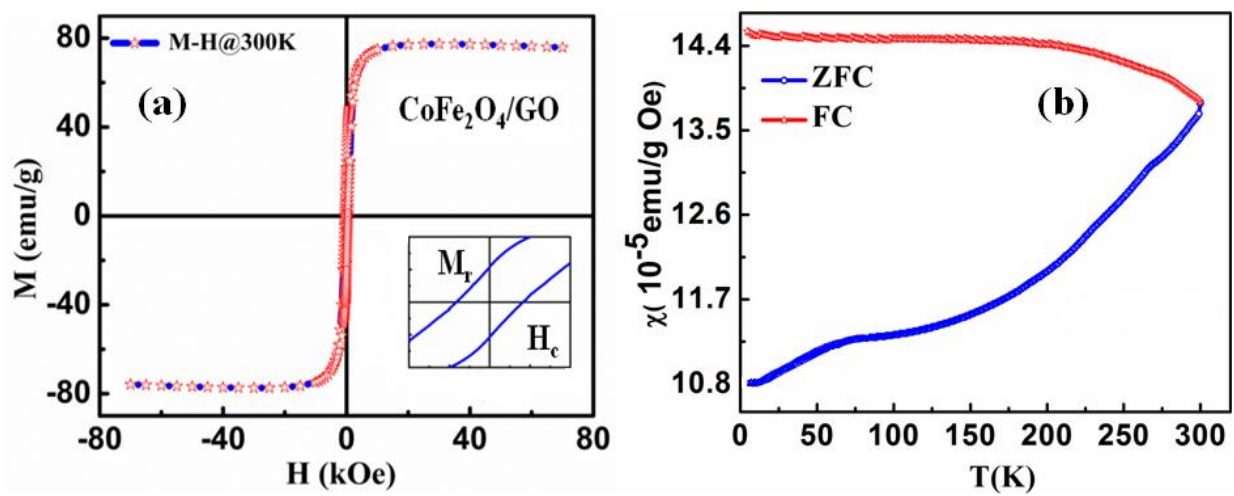

**Figure 3** (a) Hysteresis loop(M-H) of $CoFe_2O_4$/GO nanocomposite at room temperature 300 K (Inset: Zoom M-H loop of $CoFe_2O_4$/GO nanocomposite at 300 K). (b) The plot of χ vs. T of zero-field-cooling (ZFC) and field-cooling (FC) for $CoFe_2O_4$/GO nanocomposites.

The presence of superparamagnetic particles was examined using zero-field-cooling (ZFC) and field-cooling (FC) measurements with an applied magnetic field of 100 Oe. The magnetic susceptibility (χ) vs. temperature plot for $CoFe_2O_4$/GO nanocomposite is shown in Fig. 3 (b). For the zero-field-cooled (ZFC) case, the sample was cooled from 300 K to 2 K and then a magnetic field H = 100 Oe was turned on for magnetization (M) measurements with increasing temperature after ensuring stabilization at each temperature. Upon reaching 300 K, the data were similarly collected with decreasing temperature (FC mode) keeping the same applied field. It is clear that FC and ZFC curves show divergence at around 300 K which can be considered as the blocking temperature ($T_b$). Below this temperature the material shows ferromagnetic behavior above which it is superparamagnetic in nature. Above the blocking temperature, the fine nanoparticles lose their hysteresis property as evident from M–H loops.[24,25] The magnetic moments follow the direction of the applied magnetic field resulting in low remanence and low coercivity which is the characteristic feature of superparamagnetism.

**TEM Studies**

Microstructure analysis of these composites were carried out using TEM, for which in the first step, ethanol-based solutions of GO and $CoFe_2O_4$/GO were placed on carbon-coated copper grids followed by drying at room temperature before use. TEM image of GO and $CoFe_2O_4$/GO nanocomposite are shown in Fig. 4 (a) and Fig. 4 (b) respectively.

Fig. 4 (a) corresponds to the appearance of thin and wrinkled transparent GO sheets. It is consistent with our observation from the SEM analysis. Fig. 4 (b) exhibits the dispersion of $CoFe_2O_4$ NPs on GO sheets. The sheet-like corrugated morphology of GO is also well preserved in $CoFe_2O_4$/GO nanocomposite, and $CoFe_2O_4$ NPs are dispersed on GO sheet. This type of nucleation of magnetic nanoparticles on GO sheets is expected from hydrothermal synthesis. The average particles size of $CoFe_2O_4$ nanoparticles in $CoFe_2O_4$/GO nanocomposite is about $18 \pm 2$ nm. The line spacing is found to be 0.24 nm as shown in Fig. 4 (c) The selected area diffraction (SAD) pattern of $CoFe_2O_4$/GO nanocomposite is shown in Fig. 4 (d) where we observe the diffraction rings corresponding to the plane (111), (220) and (311) of $CoFe_2O_4$ as well as a diffraction pattern corresponding to hexagonally arranged carbon atoms in GO sheets.

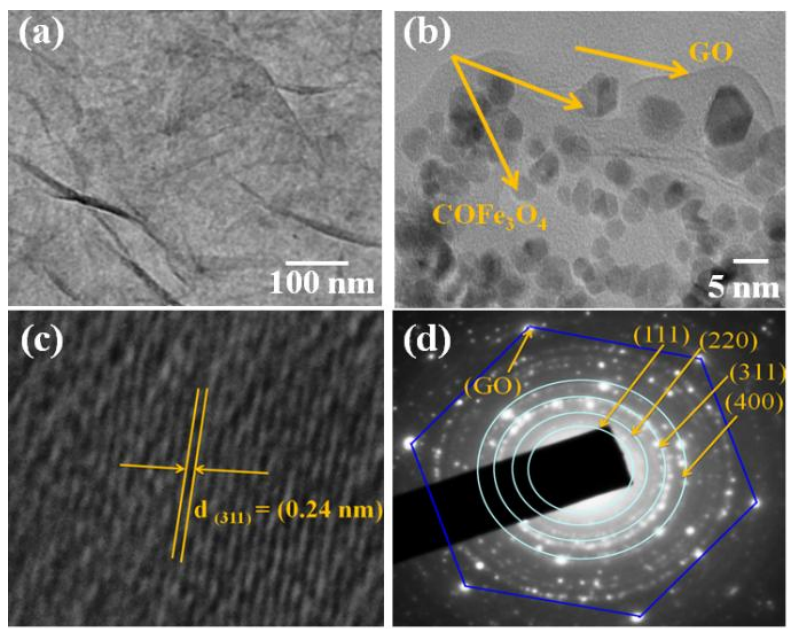

**Figure 4** TEM images of (a) GO sheets, (b) $CoFe_2O_4$/GO nanocomposite. (c) High-resolution TEM image of $CoFe_2O_4$/GO nanocomposite and (d) selected area diffraction pattern of $CoFe_2O_4$/GO nanocomposite.

**XRD measurement**

CoFe$_2$O$_4$/GO nanocomposite thin film were characterized by XRD (Panalytical 2550-PC X-ray diffractometer) set up using CuK$_\alpha$ radiation ($\lambda$=0.154nm). The data were collected between scattering angle (2θ) from 10° to 70° at scanning rate 3° min$^{-1}$. The crystalline nature of CoFe$_2$O$_4$/GO nanocomposite was identified by analyzing its X-ray diffraction (XRD) spectra as shown in Fig. 5 (a).

XRD spectrum of CoFe$_2$O$_4$/GO nanocomposite exhibits the polycrystalline nature of CoFe$_2$O$_4$ NPs having a characteristic peak at 2θ = 35.6° in addition to other peaks of CoFe$_2$O$_4$ appearing at 2θ = 18.7° (111), 30.1° (220), 35.6° (311), 43.2° (400), 54.1° (422), 57.3° (511) and 62.9° (440) (matched with JCPDS No. 75-0033). X-ray peak broadening analysis was used to calculate the crystalline sizes and lattice strain by the Williamson-Hall (W-H) analysis assuming peak widths as a function of 2θ.

The strain induced in powders due to crystal imperfection and distortion is calculated using the formula

$$\Delta\xi/\xi = \frac{\beta}{\tan\theta} \quad (1)$$

Where β is full width half maximum (FWHM) of diffraction peak (in radian) and θ is Bragg's diffraction angle (in degree) and $\Delta\xi/\xi$ is lattice strain. The crystallite size was calculated from the X-ray diffraction spectra using Scherrer's formula where, the crystallite size is inversely related to βCos θ. Considering the fact that particle size and strain are independent of each other having a Cauchy-like form, which in combination are related to FWHM by W-H equation as follows,

$$\beta\cos\theta = \frac{K\lambda}{D} + \frac{\Delta\xi}{\xi}\sin\theta \quad (2)$$

Where, the term Kλ/D represents the Scherrer's particle size distribution.[26,27] Fig. 5 (b) shows the W-H plot for CoFe$_2$O$_4$/GO nanocomposite. A linear least square fitting (5% error) to βCosθ vs. Sinθ data plot yields the value of average crystallite size (D), and lattice strain ($\Delta\xi/\xi$) to be 17 nm and 0.003 respectively. The crystallite size is the good agreement with the observed size of crystallites from TEM measurement and lattice strain is expected to be compressive type which will be discussed in forthcoming section.

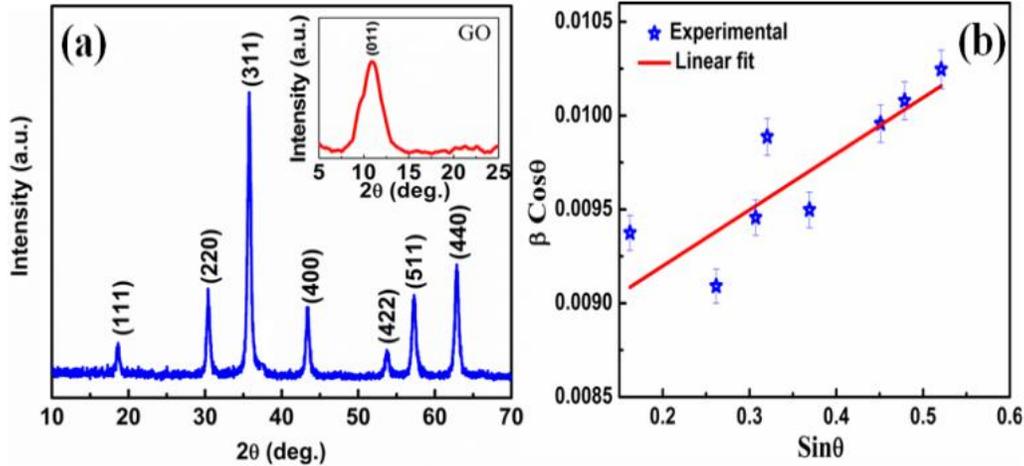

**Figure 5** (a) XRD spectrum (inset Figure for GO) and (b) Williamson-Hall plot (linearly fitted βcos θ *vs* Sinθ data) for $CoFe_2O_4$/GO nanocomposite.

**Raman Studies**

The Raman spectra of GO sheets and $CoFe_2O_4$/GO nanocomposite are shown in Fig. 6. The main features in the Raman spectra of graphene oxide sheets are D and G peaks located at 1345 cm$^{-1}$ and 1587 cm$^{-1}$ respectively. G band is attributed to the Brillouin-zone-centered LO and iTO phonon mode. D band is attributed to the double resonance excitation of phonons close to the K point scattering due to defected on iTO ($E_{2g}$) phonon in the Brillouin zone.[25,26] Spectra taken from the $CoFe_2O_4$/GO nanocomposite shows a distinct broadening of the D and G peaks of GO sheets from a full width of half-maximum (FWHM) of 122 cm$^{-1}$ to 165 cm$^{-1}$ and 69 cm$^{-1}$ to 77 cm$^{-1}$ respectively which may be due to lattice strain stemming from the interaction between GO sheets and $CoFe_2O_4$ magnetic nanoparticles.[28,29] Raman spectra of $CoFe_2O_4$/GO nanocomposite show a up shift in Raman peaks position of D and G peaks about GO sheets. The D peak is shifted from 1345 cm$^{-1}$ to 1354 cm$^{-1}$ while the G peak is shifted from 1587 cm$^{-1}$ to 1595 cm$^{-1}$ (Fig. 6). This is unlike to the reported observation of red shift in case of graphene oxide based polymer nanocomposites.[30]

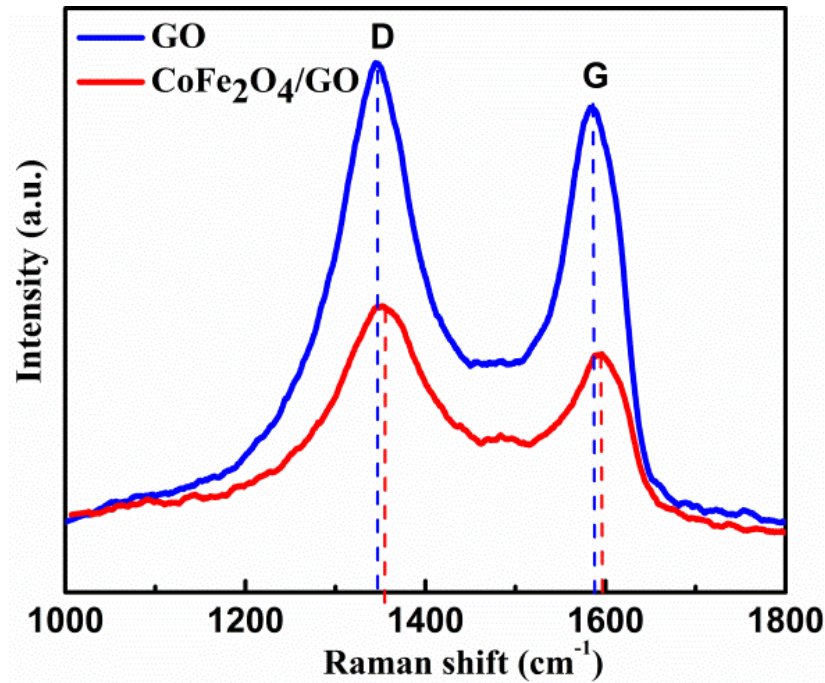

**Figure 6** Raman spectra of GO sheets and CoFe$_2$O$_4$/GO nanocomposite.

The observed shift in Raman spectrum of graphene oxide is similar to that found in graphene when subjected to lattice strain. Strain can be due to stretch in carbon-carbon bond or symmetry breaking or anisotropy in the lattice.[31,32] The direction of shift in Raman G peak is dependent on the nature of strain. It is reported that blue shift in G peak can be assigned to interfacial compressive strain. The local strain can be explained in terms of a schematic/model to understand the observed blue shift of the Raman D and G peaks in CoFe$_2$O$_4$/GO (Fig. 7). The schematics below is a depiction of TEM images where nanocomposite of CoFe$_2$O$_4$/GO are in the form of decoration of CoFe$_2$O$_4$ MNPs on GO sheets Fig.7. Lattice mismatch and disorder are expected to produce compressive stress on few layers of graphene oxide resulting in close packing of surface atoms which could have led to scattering at higher vibrational wave number.

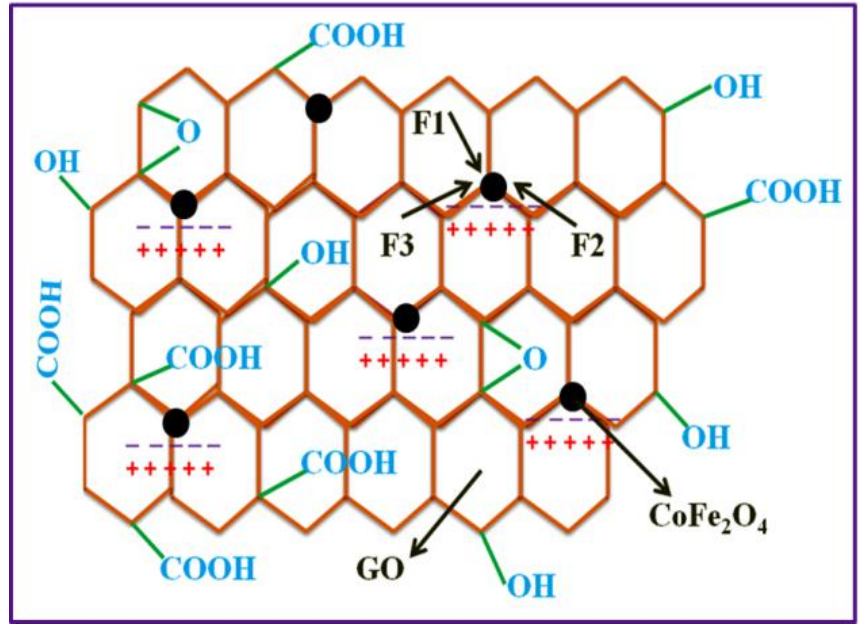

**Figure 7** Schematic illustration of interfacial compressive stress involved in CoF$_2$O$_4$/GO nanocomposite.

In Raman spectra, significant blue shift observed in D and G peaks may be attributed to simultaneous contribution from interfacial stress as well as from charge transfer process. Compressive strain involved in these vander-Waal systems could have arisen from lattice mismatch between CoFe$_2$O$_4$ nanocrystallites and GO flakes resulting in up shift in G peak. The role of defects introduced during synthesis of CoF$_2$O$_4$/GO nanocomposite also cannot be ruled out. The existence of strain is also confirmed by XRD studies as shown in Figure 5(b). In our previous result, similar type of blue shift in Raman E$_{2g}$ phonon is observed in case of Fe$_3$O$_4$/GO nanocomposites.[33]

For a hexagonal system like graphene oxide, the strain can be expressed in term of interfacial stress ($\sigma$) as,[34]

$$\omega_\sigma - \omega_0 = \alpha\sigma \qquad (3)$$

where, $\alpha = A(S_{11}+S_{12})/\omega_0$ is the stress coefficient for Raman Shift and $\sigma$ is the compressive stress. A is a constant, $S_{11}$ and $S_{12}$ are graphite elastic constants having values as $A = -1.44 \times 10^{-7}$ cm$^{-2}$, $S_{11} = 0.98 \times 10^{-12}$ Pa$^{-1}$, $S_{12} = -0.16 \times 10^{-12}$ Pa$^{-1}$ respectively.[34] $\omega_\sigma$ and $\omega_0$ are frequencies of Raman E$_{2g}$ phonon under stressed and unstressed conditions respectively. Using these constants, in equation (3), Raman shift of 8 cm$^{-1}$ in G peak of CoFe$_2$O$_4$/GO corresponds to the stress coefficient (α) and compressive stress (σ) to be 7.46 cm$^{-1}$ and 1.07 GPa respectively. We suggest that this stress might have arisen due to lattice mismatch as well as increase in defect concentration. The amount of defects present in the sample can be quantified by measuring the ratio ($I_D/I_G$) of the D and G bands. The value of $I_D/I_G$ for GO sheets and CoFe$_2$O$_4$/GO nanocomposite are found to be 1.11 to 1.30 respectively. The increased value of $I_D/I_G$ for CoFe$_2$O$_4$/GO nanocomposite as compared with GO sheets indicates the increase in disorder in GO sheets resulting from the incorporation of CoFe$_2$O$_4$ magnetic nanoparticles. The inter distance ($L_D$) between Raman active defects is estimated using Tuinstra - Koenig relation.[35,36].

$$\frac{I_D}{I_G} = \frac{C(\lambda)}{L_D}$$

Where C (λ) = (2.4 × 10$^{-10}$ nm$^{-3}$), λ$^4$ is a constant and in this case λ = 514 nm i.e the excitation wave length. The inter defect distance of GO and CoFe$_2$O$_4$/GO nanocomposite are calculated to be 15nm and 13 nm respectively. With $L_D$ >10 nm, one can expect the variation in $I_D/I_G$ ratio is due to scattered Raman active defects only. The defect density '$n_D$' is calculated using the relation[37]

$$n_D \text{ (cm}^{-1}) = 10^{14}/ \pi L^2_D$$

and is found to be 1.88x10$^{25}$ /cm$^2$ for nanocomposites thus indicating a 30% increase in point defects in GO due to nucleation of CoFe$_2$O$_4$ nanocrystallites on it.

The observed blue shift in the D and G Raman peaks and increase in FWHM confirm the occurrence of charge transfer between the GO sheets and CoFe$_2$O$_4$ NPs. The study of charge-transfer interactions of graphene with various electron donors and acceptors are reported in the literature. Charge transfer studies in CoFe$_2$O$_4$/GO and particularly its correlation with surface electronic behavior has not been reported till date. In this report, observed blue shift in G peak of GO is attributed to the situation where an electron donor molecule gets adsorbed. The increase of FWHM of G for nanocomposite band confirms the interaction with these molecules. The effect of charge transfer is quantified from shifting of Fermi surface measured by scanning Kelvin probe studies.

**SKP Studies**

In graphene oxide, Fermi level lies at Dirac point similar to that in graphene. But, in the case of $CoFe_2O_4$/GO nanocomposite where MNPs are spread over GO sheets the Fermi level of GO is expected to be changed noticeably by the charge transfer between GO and $CoFe_2O_4$ NPs. Fermi level energy of any material is related to its work function (WF) by the equation.

$$\Phi_{sample} = \chi_s + (E_C - E_F) \quad (3)$$

Where $\chi_s$ is the electron affinity of the sample, and $E_C$ and $E_F$ are the conduction band energy and Fermi energy of the material respectively.[38] The estimation of Fermi level shifting is carried out in terms of work function using scanning Kelvin probe microscopy (SKPM) setup as shown in Fig. 8. The WF value of the GO sheets is measured in terms of surface potential or contact potential difference (CPD) between GO and the reference Au tip (WF=5.1 eV). Average value of CPD is given by

$$V_{CPD} = \frac{1}{e}\left(\Phi_{tip} - \Phi_{Sample}\right) \quad (4)$$

Where $\Phi_{tip}$ and $\Phi_{sample}$ are the work function of tip and sample surface respectively and 'e' is the elementary charge on an electron.[39,40]

The measured average CPD value of GO sheet is found to be 514 mV which corresponds to WF of 4.6 eV. In the case of $CoFe_2O_4$/GO nanocomposite, obtained average CPD is -610 mV. For simplicity in plotting, the absolute value of CPD has been taken into account as shown in Fig. 8 (b). The change in contact surface potential must have emancipated from charge transfer between GO and $CoFe_2O_4$. Using equation 4, the average work function of $CoFe_2O_4$/ GO nanocomposite is found to be 5.7 eV. The significant change in the work function value of GO after decoration of $CoFe_2O_4$ nanoparticles confirms the shifting of Fermi energy level towards valence band as shown in Fig. 8 (c). This shifting may be due to electron transfer from GO to $CoFe_2O_4$ which is also envisaged from changes in Raman spectra.

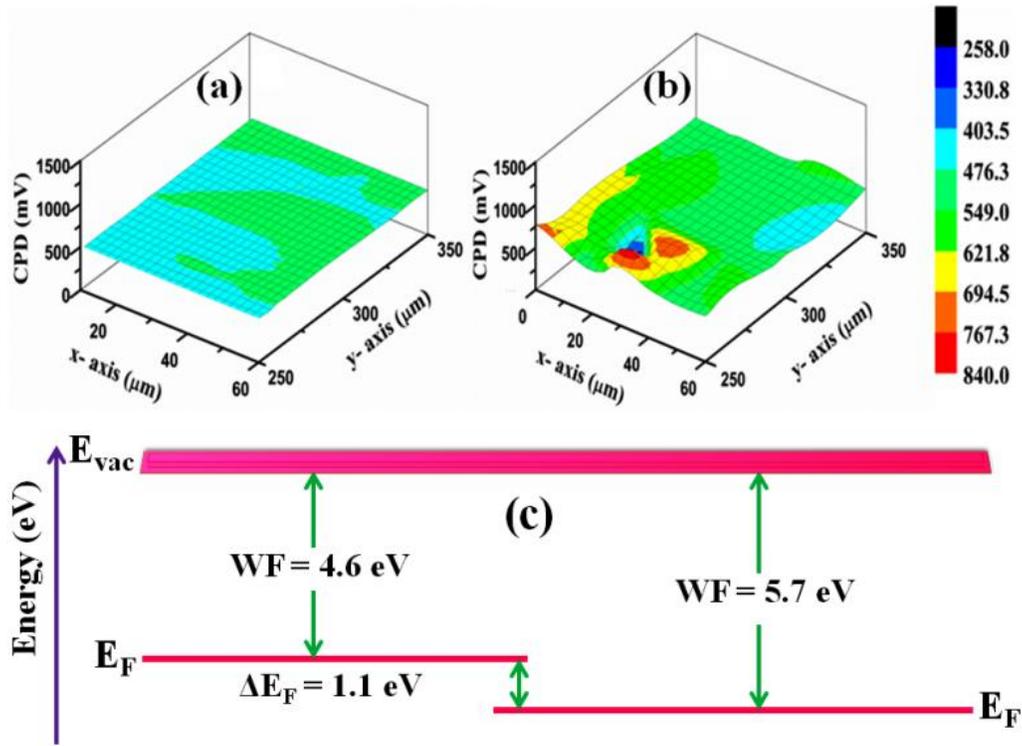

**Figure 8** CPD mapping of (a) GO and (b) $CoFe_2O_4$/GO and (c) shifting of Fermi level.

**Conclusion**

Dispersion of $CoFe_2O_4$ nanoparticles on GO sheets involves interfaical compressive stress as well as charge transfer between host GO sheets and $CoFe_2O_4$ magnetic nanoparticles. The superparamagnetic behavior of these nanocomposite is confirmed from its high value of magnetic saturation with Ms (75.37 emu/g) and low coercivity value with $H_c$ (0.41 kOe), thus indicating soft magnetic nature of $CoFe_2O_4$/GO nanocomposite. Charge transfer process induces a blue shift in $E_{2g}$ phonon as well as an increase in FWHM of Raman spectra of GO sheets. Compressive strain calculated from XRD peak is related to the observed blue shift in Raman peak. Point defects generated in these nanocomposites are of the order of $10^{25}$ per cc which play an important role for generation of interfacial compressive stress as well as charge transfer process. The effect of charge transfer is quantified in terms of changes in surface potential of GO, leading to a shift in Fermi surface towards valence band.

**Acknowledgments**

The authors are thankful to AIRF, JNU, New Delhi for providing XRD, SEM, characterization. Amodini Mishra is thankful to UGC for providing fellowship.